\documentclass[11pt]{article}
\usepackage[utf8]{inputenc}
\usepackage{gensymb}
\usepackage{adjustbox}
\usepackage{booktabs}
\usepackage{multirow}
\usepackage{longtable, tabularx}
\usepackage{epsfig}
\usepackage{bm}
\usepackage{lscape}
\usepackage{mathtools}
\usepackage{amsmath, amssymb}
\usepackage{array,longtable}
\usepackage{xcolor}
\usepackage{hyperref}
\usepackage{graphicx}

\let\code=\texttt
\usepackage{ctable}
\usepackage{algorithm}
\usepackage{algorithmic}
\usepackage[round,sort]{natbib}

\newcommand{\argmax}{\arg\!\max}

\newcommand*{\tran}{^{\mkern-1.5mu\mathsf{T}}}

\usepackage[utf8]{inputenc}
\usepackage{scalerel,stackengine}
\stackMath
\newcommand\reallywidehat[1]{%
\savestack{\tmpbox}{\stretchto{%
\scaleto{%
\scalerel*[\widthof{\ensuremath{#1}}]{\kern.1pt\mathchar"0362\kern.1pt}%
{\rule{0ex}{\textheight}}
}{\textheight}%
}{2.4ex}}%
\stackon[-6.9pt]{#1}{\tmpbox}%
}
\newcommand{\comment}[1]{}
\usepackage{caption}
\usepackage{caption}
\addtocontents{file}{text}
\usepackage{geometry}
\usepackage{soul}

\definecolor{eGreen}{rgb}{.057, .549,.065}
\newcommand{\thad}[1]{{\textcolor{blue}{#1}}}

\makeatletter
\patchcmd{\maketitle}{\@fnsymbol}{\@alph}{}{}  
\makeatother

\title{Extracting Scalar Measures from Curves}

\author{%
  \begin{minipage}{.45\textwidth}
    \centering
    \href{https://orcid.org/0000-0001-5935-6018}{\includegraphics[scale=0.06]{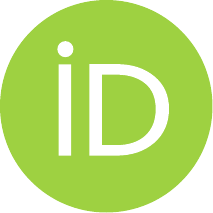}\hspace{1mm}Lanqiu Yao} \\
    Magma Capital Funds\\
    233 S Wacker Dr \#4400 \\
    Chicago, IL 60606\\
    \texttt{lyao@magmacapitalfunds.com}
  \end{minipage}%
  \hspace{.1\textwidth}
  \begin{minipage}{.45\textwidth}
    \centering
    \href{https://orcid.org/0000-0002-3964-5899}{\includegraphics[scale=0.06]{orcid.pdf}\hspace{1mm}Thaddeus Tarpey} \\
    Division of Biostatistics, \\ 
    Department of Population Health,\\
    NYU Grossman School of Medicine \\
    New York, NY 10016 \\
    \texttt{Thaddeus.Tarpey@nyulangone.org}
  \end{minipage}
}
\date{\vspace{-5ex}}
                  
\begin{document}
             
\maketitle

{\bf Abstract.}
The ability to order outcomes is necessary to make comparisons which is complicated when there is no natural ordering on the space of outcomes, as in the case of functional outcomes.
This paper examines methods for extracting a scalar summary from functional or longitudinal outcomes based on an average rate of change which can be used to compare curves.
Common approaches used in practice use a change score or an analysis of covariance (ANCOVA) to make comparisons.  
However, these standard approaches only use a fraction of the available data and are inefficient.  We derive measures of performance of an averaged rate of change of a functional outcome and compare this measure to standard measures. 
Simulations and data from a depression clinical trial are used to illustrate results.

\section{Introduction} \label{introduction}

The most frequently used inference procedure in practice is arguably that of comparing two or more groups, such as treatment versus control, in terms of specific outcome measures. 
It is common to collect outcome measures at multiple timepoints, with the primary focus being the {\em change} over time in the outcome of interest.  
The simplest and perhaps most frequently used measure of change is the change score (e.g., last $-$ baseline).  
However, this simple change score is inefficient in longitudinal settings where the outcome is measured multiple times, as it only uses the first and last measurements, thus ignoring most of the data. 
The analysis of covariance (ANCOVA) approach, which regresses the last (or the change score) on the baseline along with an indicator that distinguishes groups is similarly inefficient.
If data for individuals is collected over some domain (assumed to be time in this paper), such as functional data \citep{fda2}, then comparing groups for outcomes in such domains is complicated because there is typically no natural ordering on curves and surfaces, especially if one-sided inference is of interest.
In the realm of precision medicine, many individualized treatment rule (ITR) approaches have been developed for scalar outcomes \cite[e.g.,][]{Ciarleglio.2015, ParkEtAl2021}. 
However, very few have investigated the development of ITRs that take full advantage of the efficiencies offered by using the complete functional nature of the data \citep[see][]{yao2022single}.

The focus of this paper is to investigate and compare approaches for extracting scalar summaries from functional outcomes, aiming to establish an ordering that enables more powerful inferential comparisons.
The change score divided by the length of the time interval (denoted as CS) is a crude estimator of the rate of change. 
We consider extracting a refined summary measure from a curve based on a weighted average of the rate of change across the entire domain for the curves.  
Because the instantaneous rate of change at any given time point is represented by the tangent slope (calculated from the first derivative), we have named the corresponding summary measure the {\em weighted average tangent slope} (WATS), which will be defined in Section \ref{watssection}. 
Estimators of the WATS utilize all available data and consequently have far superior performance than approaches based on CS and ANCOVA.
When a uniform weight function is applied to WATS, the resultant summary measure is termed the Average Tangent Slope (ATS).
In Section \ref{sectionats}, we introduce a parametric estimator of ATS, referred to as {\em mean change} (MC) estimator, and conduct a comparison with the CS.

The problem of extracting summary measures from functional outcomes has a long history \cite[e.g.,][]{wishart1938growth}.  
The critical process of these methods is to extract an appropriate summary statistic from the curves. 
Other common summary measures for curves, in addition to the CS, include the mean, maximum, or the value of the final outcome. 
Another natural summary measure to use when the functional outcomes are straight lines is simply the slope of the line \citep{vossoughi2012summary}.  
However, the model of straight lines often does not fit the observed data well unless the time domain is very short. 
ANCOVA, which conditions on baseline values, is a popular alternative inferential approach.
The debate over whether to use ANCOVA or CS as a summary measure has been extensive \citep{bland2011comparisons, senn2000repeated,senn2006change,vollmer1988comparing}.
For example, \cite{lord1967paradox} highlighted that the CS and ANCOVA (adjusted for baseline) can yield divergent conclusions, which is known as Lord's paradox. 
Scenarios have been described to demonstrate that ANCOVA is superior in terms of bias and efficiency \citep{van2006ancova, liu2009should}. 
Conversely, the CS may outperform ANCOVA in certain scenarios, such as when the variance between subjects is greater than the variance within subjects \citep[e.g.,][]{norman1989issues}. 
Other methodologies, such as linear mixed-effect (LME) models and generalized estimating equations (GEE), are also prevalent in longitudinal data analysis. These methods are robust and make full use of the data, yet they do not inherently generate a singular scalar summary of outcome trajectories to facilitate treatment comparisons or the derivation of ITRs.

The ongoing discussion over the optimal choice of summary statistics for curves in longitudinal studies encounters additional complexity due to the prevalent issue of missing data.
This common challenge, such as participant dropout, substantially impacts study outcomes \citep{hong2018generalizing}. 
The presence of missing data can significantly distort the estimations derived from CS, underscoring the need for more robust analytical methods.
To tackle this issue, this paper compares various methods for estimating the change of the outcomes in the context of missing data, providing insights into effective approaches under these circumstances.
In this context, we introduce the MC estimator, a novel parametric approach tailored for extracting scalar summaries from functional outcomes.
Despite its intuitive appeal and suitability for the task,
the MC estimator has received little mention in the literature.
Its principal strength lies in the comprehensive utilization of all available data, which enhances its analytical robustness. Additionally, when derived from a model fitted using maximum likelihood methods, the MC estimator can be expected to benefit from the usual efficiencies enjoyed by maximum likelihood estimators.
One concern about the MC estimator is that it depends strongly on the proposed model used to fit the data.  However, most longitudinal randomized studies are typically of short time duration (e.g., 6-12 weeks) in which case relatively simple functional forms (e.g., quadratic curves) often provide very good approximations to the outcome trajectories. 
For example,
it is common in many randomized clinical trials (RCTs) for individuals to experience either an immediate response due to placebo effects and/or immediate specific treatment effects whose effects may diminish or strengthen over time. 
In these settings,
straight line models provide poor fits to the data whereas quadratic model can usually capture these response characteristics.
For studies of longer duration, cubic $B$-splines (often with a single knot point) will suffice to provide sufficient flexibility to obtain good fits to individual outcome trajectories in which case the MC estimator may be expected to perform well.  
If additional flexibility is needed, nonparametric smoothing methods can be used to fit the curves; however the focus of this paper is on parametric models typically used in randomized studies.

The rest of the paper is organized as follows: 
Section \ref{watssection} introduces the definitions and notation used throughout the paper; 
Section \ref{sectionats} offers a detailed comparison of the MC and CS estimators; Section \ref{hyptest2} is dedicated to contrasting the CS, MC, and ANCOVA, as well as the slope derived from fitting a straight line, in the context of hypothesis testing; 
Section \ref{watsection} describes a method for estimating a non-uniform weight function in the WATS; 
an illustrative example from a depression RCT is presented in Section \ref{embarc}; 
and the paper concludes with Section \ref{discussion}.

\section{The Weighted Average Tangent Slope (WATS)}\label{watssection}

Let $y_{ik}(t)$ represent a continuous and smooth functional outcome, such as a longitudinal trajectory, for subject $i$ within the $k$th group, where $i \in \{1, \ldots, n_k\}$, and $k \in \{1, \ldots, K\}$.
For the sake of clarity, we assume that the argument $t$ of these functions denotes time throughout.
We can express $y_{ik}(t)$ as the sum of two components: $\mu_k(t)$, which represents the mean function of the $k$th group, and $h_{ik}(t)$, a subject-specific random effect with $E(h_{ik}(t)) = 0$.
The observed outcomes are denoted as $\tilde{y}_{ijk}$, where $j \in \{1,..., m_{ik}\}$ signifies the index of individual observations. 
We represent the vector of observed outcomes along with their corresponding observation times as $\tilde{\bm Y}_{ik} = (\tilde{y}_{i1k},  \ldots, \tilde{y}_{im_{ik}k})\tran$ and $\bm t_{ik} = (t_{i1k}, t_{i2k}, \ldots, t_{im_{ik}k})\tran$, respectively.
Consequently, we can formulate the relationship as follows:
\begin{equation}\label{01}
    \tilde{y}_{ijk} = \mu_k(t_j) + h_{ik}(t_j) + \epsilon_{ijk},
\end{equation}
where
$\epsilon_{ijk}$ 
is a random error (noise).
Typically, research studies are designed to maintain balance by adhering to a common set of follow-up times, denoted as $t_{ijk} = t^*_j$, where each subject contributes a total of $m$ observed outcomes. 
In this paper, we also explore the implications of subjects having missing observations.
Drawing from the assumption of smoothness, we can express $\mu_{k}(t)$ as $\bm g(t) \tran \bm \beta_k$ and $h_{ik}(t)$ as $\bm g(t) \tran \bm b_{ik}$. Here, $\bm g(t) = [g_1(t), \ldots, g_p(t)] \tran$ represents a set of basis functions, while $\bm \beta_k$ and $\bm b_{ik} \in \mathbb{R}^p$ correspond to the coefficients associated with fixed and random effects, respectively.
We assume $\bm b_{ik}$ follows a multivariate distribution with mean equals 0 and covariance matrix $\bm D_k$.
This formulation allows us to represent the relationship as follows:
\begin{equation}\label{02}
\tilde{\bm Y}_{ik} = \bm X_{ik} (\bm \beta_k + \bm b_{ik}) + \bm \epsilon_{ik},
\end{equation}
where $\bm X_{ik} = [\bm g(t_{i1k}), \ldots, \bm g(t_{im_{ik}k})] \tran$ is the design matrix for the mixed effect model. 
Let $\bm X^* = [\bm g(t_1^*), ..., \bm g(t_m^*)]\tran$ denote the design matrix when outcomes are obtained at all design timepoints, i.e., no missing outcomes exist.
The model can be generalized to incorporate other baseline covariates, e.g., age, gender, etc. 
Additionally, we assume that an adequately rich set of basis functions is used in the model ($\ref{02}$) so that any model misspecification is minimal and the error  $\epsilon_{ijk}$ has essentially a mean zero across timepoints.

In order to extract a meaningful scalar summary from a curve, it is natural to consider an overall summary of the rate of change.
Given that the mean instantaneous rate of change at time $t$ for the mean trajectory is $\mu_k^\prime (t)$, as established in (\ref{01}), a scalar summary measure for a curve, as proposed by \cite{tarpey2021extracting}, can be defined by employing a weight function $w(t)$ tailored for functional outcomes, as discussed in \cite{chen2014optimally}. 
This definition takes the form of:
\begin{equation}\label{wats0}
    \textbf{Weighted Average Tangent Slope (WATS):= } \text{ }
    \int_{t_1^*}^{t_m^*} w(t) \mu_k^\prime (t) dt
    =    \int_{t_1^*}^{t_m^*} w(t) \bm g^\prime(t)\tran \bm \beta_k dt = \bm S_w \tran \bm \beta_k.
\end{equation}
In this expression, $\bm S_w = \Big(
\int_{t_1^*}^{t_m^*} w(t) g_1^\prime(t)dt, \ldots, 
\int_{t_1^*}^{t_m^*} w(t) g_p^\prime(t)dt\Big) \tran$ is a vector incorporating components obtained by integrating the weighted derivatives of basis functions.
The weight function is designed to meet the conditions: $w(t) \ge 0$ and $\int_{t_1^*}^{t_m^*} w(t) dt = 1$, where $t_1^*$ and $t_m^*$ denote the initial and final timepoints for the study.

Often, the question of primary importance regards the overall average change in the outcome over time, as defined by:
\begin{equation}\label{diff}
{\mu_k(t_m^*) - \mu_k(t_1^*)\over t_m^*-t_1^*}.
\end{equation}
When we apply a uniform weight function to calculate the WATS (\ref{wats0}), 
where $w(t) =  1/(t_m^*-t_1^*)$ for $t_1^*< t < {t_m^*}$,
the resulting WATS is equivalent to  (\ref{diff}).
In such cases, we refer to the WATS with a uniform weight function as simply the {\em Average Tangent Slope} (ATS).
Utilizing
$\bm G = \frac{\bm g(t^*_m) - \bm  g(t^*_1)}{t_m^* - t_1^*} \in \mathbb{R}^p$ from (\ref{wats0}),
we have
\begin{equation}\label{ats0}
  \textbf{Average Tangent Slope (ATS):= }  \text{ }
 \text{ } \frac{1}{t_m^*-t_1^*} \int_{t_1^*}^{t_m^*} \mu_k^\prime (t) dt= \frac{\mu_k(t_m^*) - \mu_k(t_1^*)}{t_m^*-t_1^*} = 
     \bm G\tran {\bm \beta}_k.
\end{equation}
In cases where the trajectories are linear, the ATS corresponds to the average slope of these lines across individuals.

\section{Comparison of Change Score and Mean Change Estimators}
\label{sectionats}

A straightforward nonparametric approach for estimating the ATS is to employ the  {\em Change Score} (CS) estimator, which is defined as:
 \begin{equation} \label{CS}
\reallywidehat{\text{CS}}_k =
\frac{1}{n_k}\sum_{i=1}^{n_k} \frac{ \tilde{y}_{im_{ik}k} - \tilde{y}_{i1k}}{t_{im_{ik}} - t_{i1}}.
\end{equation}
While the CS estimator is conceptually simple and intuitive, it has a limitation in that it only considers the first and last observations, making it inefficient as it disregards potentially valuable information from other observations.
A more efficient approach to estimate the ATS is by utilizing a parametric estimator for ${\bm \beta}$ in (\ref{ats0}), resulting in the \textit{Mean Change (MC)} estimator, defined as:
\begin{equation} \label{ats1}
\reallywidehat{\text{MC}}_k = \frac{\reallywidehat{\mu}_k(t_m^*) - \reallywidehat{\mu}_k(t_1^*)}{t_m^*-t_1^*} =
\bm G\tran \reallywidehat{\bm \beta}_k,
\end{equation}
where $\reallywidehat{\bm \beta}_k$ represents the estimated fixed-effect vector derived from model (\ref{02}), which is typically obtained using maximum likelihood estimation (MLE). 
The key advantage of the MC lies in its ability to leverage all available data, thereby achieving increased efficiency compared to the CS. 
Furthermore, the MC is more robust in the presence of challenges arising from missing data.

Following (\ref{ats1}), the MC is unbiased for the ATS if the estimator $\reallywidehat{\bm \beta}_k$ is unbiased.
For instance, the MC is (approximately) unbiased when estimating $\bm{\beta_k}$ in (\ref{ats1}) using maximum likelihood, and when the missingness of outcome data follows the missing at random (MAR) assumption.
In contrast, the CS is an unbiased estimator for the ATS only when there is no missing data. 
However, when data is missing, the CS can become biased, particularly if there is a probability of the last observation being missing when the $\mu_k(t)$ from (\ref{01}) is not constant.
To illustrate this, we introduce a random variable $T$ to denote the last observation time point for an individual. Assume that $T$ can take on values such as $t_{j}^*$ with probabilities denoted by $p_{j}$, subject to the constraint that $\sum_{j = 2}^m p_{j} = 1$  (where it is assumed that the baseline outcome at time $t_{1}$ is always observed).
Under this scenario, the expectation of CS estimator can be expressed as: 
\begin{equation}\label{csbias1}
     E(\reallywidehat{\text{CS}}_k) =  E\big(E(\reallywidehat{\text{CS}}_k | T) \big)
     = \sum_{j=2}^m p_j E(\reallywidehat{\text{CS}}_k | T = t_{j}^*)
     =  \sum_{j=2}^m {p_j d_j},
\end{equation}
where $d_j=(\mu_k(t_j^*) - \mu_k(t_1^*))/(t_j^* - t_1^*).$ In order for $\reallywidehat{\text{CS}}_k$ to be an unbiased estimator of the ATS, from (\ref{csbias1}), it must hold that
$ \sum_{j=2}^m {p_j d_j} = d_m$ for the family of discrete probability distributions on $\{t_2^*,\dots, t_m^*\}$. Noting $p_2+\cdots + p_m=1$, unbiasedness of 
$\reallywidehat{\text{CS}}_k$ would imply $\sum_{j=2}^{m-1}{p_j(d_j-d_m)}=0$ uniformly across the parameter space defined by the $p_j$. 
However,  this cannot hold in general since a system of linear equations identically equal to zero requires that all the coefficients are zero (in this setting, this would require $d_j-d_m=0$ for
$j=2,\dots, m-1$ which does not hold in general when $\mu_k(t)$ is non-constant).

The main result in this section is given by the following proposition:

\textbf{\textit{Proposition 1:}} Under the usual linear mixed-effect model assumptions of normality and independence of the random effects and the error in model (\ref{02}), 
the MC estimator of the ATS
derived from normal theory maximum likelihood estimation is more efficient than the CS estimator, i.e.,
$\text{Var}(\reallywidehat{\text{MC}}_k)  \leq \text{Var}(\reallywidehat{\text{CS}}_k)$, when there is no missing data.
\medskip

Although this result aligns with intuition and holds asymptotically due to the large sample optimality of maximum likelihood estimation, we have found no formal derivation of this result in the literature without resorting to asymptotic justification.
The proof of Proposition 1 follows from first noting that $\mbox{Cov}(\tilde {\bm Y}_{ik}) =
{\bm V}_{ik} = \bm X_{ik} {\bm D}_k \bm X_{ik}\tran + {\sigma}_k^2 \bm I $ and
the estimators of the $\bm \beta_k$ are obtained from 
$(\sum_{i=1}^n \bm X_{ik}\tran  {\bm V}_{ik}^{-1} \bm X_{ik})^{-1} (\sum_{i=1}^n \bm X_{ik}\tran  {\bm V}_{ik}^{-1} \tilde{\bm Y}_{ik})$,
where ${\bm D}_k$ is the covariance matrix for the random effects and $\sigma^2_k$ is the variance of the error term
\citep[e.g., see][]{hedeker2006longitudinal}.
The variance of the CS estimator is
\begin{equation} \label{var_cs0}
\begin{aligned}
 \text{Var} (\reallywidehat{\text{CS}}_k) 
      =   \text{Var} \Big(\frac{1}{n_k}\sum_{i=1}^{n_k} \bm h_{ik} \tran \tilde {\bm Y}_{ik}\Big) 
      =  \frac{1}{n_k} \bm h\tran \bm V_k \bm h,
 \end{aligned}
\end{equation}
where $\bm h = (\frac{-1}{t_m^* - t_1^*}, 0,..., 0,\frac{1}{t_m^* - t_1^*})\tran \in \mathbb{R}^m$.

From (\ref{ats1}), 
$\text{Var}(\reallywidehat{\text{MC}}_k)  = \bm G\tran\text{Cov} ( \reallywidehat{\bm \beta}_k) \bm G$,
where the covariance matrix of $\reallywidehat{\bm \beta}_k$ is 
\begin{equation} \label{13}
    \text{Cov} ( \reallywidehat{\bm \beta}_k)  = \text{Cov} \Big( (\sum_{i=1}^{n_k} \bm X_{ik}\tran  {\bm V}_{ik}^{-1} \bm X_{ik})^{-1} (\sum_{i=1}^{n_k} \bm X_{ik}\tran  {\bm V}_{ik}^{-1} \tilde{\bm Y}_{ik}) \Big) 
  = \frac{1}{n_k} (\bm X^{* \mkern-1.5mu\mathsf{T}} {\bm V_k}^{-1}  \bm X^*)^{-1}. 
\end{equation}
Applying the Woodbury matrix identity \citep{max1950woodbury}
$\big(\bm X_{ik}\tran  {\bm V}_{ik}^{-1} \bm X_{ik}\big)^{-1} =  {\sigma}_k^2 (\bm X_{ik}\tran \bm X_{ik})^{-1} + {\bm D}_k$
and noting that
$    \bm h \tran \bm X^* = \frac{\bm g(t_m^*) \tran - \bm  g(t_1^*)\tran}{t_m^* - t_1^*} = \bm G \tran$,
from (\ref{var_cs0}), we can write
\begin{equation}\label{varCS1}
    \begin{aligned}
        \text{Var}( \reallywidehat{\text{MC}}_k ) &=   \frac{1}{n_k}\bm G \tran  \big({\sigma_k^2} (\bm X^{* \mkern-1.5mu\mathsf{T}} \bm X^*)^{-1} + \bm D_k \big) \bm G \\
        & =  \frac{1}{n_k} \bm h \tran \bm V_k \bm h + \frac{\sigma_k^2}{n_k}\bm h \tran \bm X^* (\bm X ^{* \mkern-1.5mu\mathsf{T}} \bm X^* )^{-1}  \bm X ^{* \mkern-1.5mu\mathsf{T}} \bm h - \frac{\sigma_k^2}{n_k}\bm h \tran \bm h\\
   & = 
     \text{Var} (\reallywidehat{\text{CS}}_k) -
     \frac{\sigma_k^2}{n_k} (\bm h \tran \bm h)
     \Bigl (1-{{\bm h \tran \bm X^* (\bm X ^{* \mkern-1.5mu\mathsf{T}} \bm X^* )^{-1}  \bm X ^{* \mkern-1.5mu\mathsf{T}} \bm h}\over  \bm h \tran \bm h} \Bigr )\\
     &   \le 
     \text{Var} (\reallywidehat{\text{CS}}_k)\\
    \end{aligned}
\end{equation}
where the last inequality (\ref{varCS1}) follows because $ \bm h \tran \bm h > 0$ and
the matrix
$\bm X^* (\bm X ^{* \mkern-1.5mu\mathsf{T}} \bm X^* )^{-1}  \bm X ^{* \mkern-1.5mu\mathsf{T}}$  is idempotent whose eigenvalues are $0$'s or $1$'s and therefore
the maximum value of the Rayleigh Quotient 
${{\bm h \tran \bm X^* (\bm X ^{* \mkern-1.5mu\mathsf{T}} \bm X^* )^{-1}  \bm X ^{* \mkern-1.5mu\mathsf{T}} \bm h}\over  \bm h \tran \bm h}$
is $1$.
Note that the inequality in (\ref{varCS1})  is strict unless $\bm{h}$ is proportional to an eigenvector of 
$\bm X^* (\bm X ^{* \mkern-1.5mu\mathsf{T}} \bm X^* )^{-1}  \bm X ^{* \mkern-1.5mu\mathsf{T}}$ corresponding to an eigenvalue of 1.
Since $\bm h = (\frac{-1}{t_m^* - t_1^*}, 0,..., 0,\frac{1}{t_m^* - t_1^*})\tran$ has all zero entries except for the first and last entry, $\bm{h}$ 
will typically not correspond to a multiple of an eigenvector and therefore the inequality in (\ref{varCS1}) will be strict. Furthermore, it is important to note a subtle yet significant aspect of our analysis: in practical applications, the estimation of parameters such as ${\bm D}_k$ and $\sigma_k^2$ is necessary to obtain $\reallywidehat{\bm \beta}_k$ and subsequently the estimator of the MC. However, the current proof does not encompass the estimation of these parameters. Instead, what is demonstrated is $ \text{Var}(\reallywidehat{\text{MC}}_k) \leq \reallywidehat{\text{Var}}(\reallywidehat{\text{CS}}_k)$, rather than $\reallywidehat{\text{Var}}(\reallywidehat{\text{MC}}_k) \leq \reallywidehat{\text{Var}}(\reallywidehat{\text{CS}}_k)$. This distinction is crucial for the interpretation of our results and should be carefully considered in practical implementations.

When there is missing data, the variance of the MC increases (details not shown).  
However, it is fairly easy to concoct examples where the variance of the CS can either increase or decrease when there is missing data for the final observation. 
However, depending on the shape of the trajectories, missing final observations can severely bias the estimation of the ATS using the CS.

\section{Hypothesis Testing} \label{hyptest2}

A primary motivation and major advantage of extracting the ATS scalar summary is that it provides an inferential means for formally comparing groups when outcomes are curves. 
This is particularly valuable for one-sided inference, a scenario common when dealing with curves since they often lack a formal ordering. 
In this paper, our emphasis is on evaluating the differences between groups; therefore, a two-sample test framework was considered.
Since the CS is estimated from sample means, a simple 2-sample $t$-test can be employed to compare groups using the CS.  
For the MC, one obtains estimates $ \reallywidehat{\text{MC}}$  and $\reallywidehat{\text{cov}}(\reallywidehat{\bm \beta}_1)$ from the fitted model (\ref{ats1}). 
This enables the computation of a Wald test statistic to facilitate group comparisons:
\begin{equation}\label{wald}
W^{\text{MC}} = \frac{\reallywidehat{\text{MC}}_1- \reallywidehat{\text{MC}}_2}{\sqrt{ \reallywidehat{\text{Var}} (\reallywidehat{\text{MC}}_1) +  \reallywidehat{\text{Var}} (\reallywidehat{\text{MC}}_2)}} =
 \frac{\reallywidehat{\text{MC}}_1 - \reallywidehat{\text{MC}}_2}{\sqrt{ \bm G\tran (\reallywidehat{\text{cov}}(\reallywidehat{\bm \beta}_1)+\reallywidehat{\text{cov}}(\reallywidehat{\bm \beta}_2)) \bm G}}.
\end{equation}
Additionally, ANCOVA is another prevalent method for group comparison when dealing with curve outcomes. This approach, similar to the CS, leverages only the baseline and final observation in its estimator model:
$$\text{Last Observation} = \alpha_0 + \alpha_1 \text{Baseline} + \alpha_2 \text{Group} + \text{Error}.$$
Hypothesis testing in ANCOVA typically involves assessing  $H_0: \alpha_2 = 0$ using a $t$-test statistic.  
The subsequent section presents simulation results to evaluate and compare the testing performances of the CS, MC, and ANCOVA approaches.

\subsection{Simulation Illustrations} \label{simulation1}

This section summarizes results from extensive simulations that compare the three approaches — CS, MC, and ANCOVA — in terms of statistical power and control of Type I error rates across a broad range of scenarios.

\subsection{Simulation Settings}\label{settings}

\begin{figure}[htp]
\caption{\textbf{Simulated Outcome Trajectories}}
\centering
\includegraphics[width=0.9\textwidth]{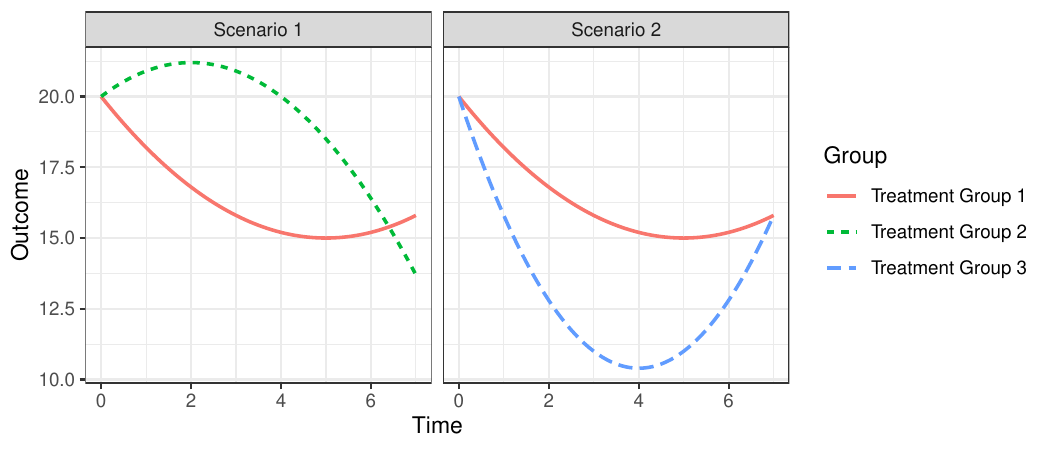}
\label{f0}
\caption*{
\footnotesize{
The mean trajectories of groups are shown for two simulation scenarios: 
Scenario 1: the two mean curves have different shapes and different ATSs; 
Scenario 2: the mean two curves have different shapes but the same ATS. 
For Scenario 3 (not shown), the two mean trajectories are identical for both groups (e.g., represented by the red parabola). 
}}
\end{figure}

{\bf Mean Trajectory Settings:}
In the simulation study, we focus on comparing outcome trajectories between two groups. 
Quadratic curves, similar to what is often seen in longitudinal depression studies \citep[e.g.,][]{tarpey2021extracting}, were simulated using the model described in (\ref{02}). 
Additionally, we considered non-quadratic scenarios to broaden our analysis scope. 
Details on the simulation settings and results of the non-quadratic scenario are provided in the Appendix.
We considered three distinct mean trajectory scenarios for these simulations:
(i) ATS differs for the two groups, 
(ii) ATS is identical for the two groups, but their mean trajectories differ, and 
(iii) mean trajectories are identical for both groups.
Figure \ref{f0} illustrates the mean trajectories for the first two scenarios, while the mean trajectory for the third scenario (identical curves for both groups) is represented by the solid red parabola in the same figure.
The chosen outcome time points were set at $t \in \{0, 1, ..., 7\}$, mirroring common practice in clinical trials.
Details on parameter settings for these scenarios are provided in the Appendix.
Note that across all scenarios, the mean outcome value at baseline is consistently identical for both groups. 
This uniformity aligns with the expectations set by randomized experiments. 
Specifically, Scenarios 2 and 3 are designed to represent the null hypothesis, which posits that the ATS remains equivalent for both groups. 
However, a notable distinction arises in Scenario 2: despite both groups having the same ATS, their mean trajectories exhibit differences. 
This contrast highlights the unique characteristic of Scenario 2, where distinct trajectories can lead to the same ATS.

{\bf Missing data settings:}
We considered four distinct scenarios of missing data:
(i) No missing observations: All subjects had complete data records without any missing observations.
(ii) Missing completely at random (MCAR): Each subject had a 15\% probability of missing outcomes at random for any observation time point, excluding the baseline. 
This led to an average of 15\% missing data points within each participant's dataset, randomly scattered across the time points, with baseline data intact.
(iii) Dropout-induced missingness: This scenario modeled a monotone pattern of missingness due to dropout, independent of the outcome. 
Specifically, the distribution of each subject was set as follows: 50\% with no missing data, 30\% missing the final observation, and 10\%, 5\%, and 5\% missing the last two, three, and four observations, respectively.
(iv) Missing not at random (MNAR): A latent (unobserved) variable $z_{ijk} \sim N(0, 3^2)$ was included in (\ref{02}), that is, 
   $ \tilde{\bm Y}_{ik} = \bm X_{ik} (\bm \beta_k + \bm b_{ik}) + \bm z_{ik} +\bm \epsilon_{ik},$
where outcomes with $z_{ijk} < -1.15$ were set to missing (except for the baseline).
To address these missing data scenarios, we employed Complete Record Analysis (CRA) and Multiple Imputation (MI) approaches \cite{rubin1976inference}.

{\bf Error variance:} 
Data was generated with a range of noise variances.
The standard deviations for the random error
$\epsilon_{ijk}$ in (\ref{01}) were set at \{0.5, 1, 1.5, 2, 2.5, 3\}.

In summary, the simulation study evaluated the following combinations of settings: 
three mean trajectory scenarios,
six levels of error variances, 
and four missing data mechanisms, 
culminating in a total of 72 distinct scenarios $(3 \times 6 \times 4 = 72)$.
For each scenario, a sample size of $n_k=100$ was generated for each group.
The approaches used for ATS estimation included the MC, CS, and the slope of a linear fit. 
Mixed-effect models were estimated using the R package \code{lme4}. 
For scenarios with missing data, MI was conducted using the \code{mice} package \citep{mmm}.

\subsection{Simulation Results}

In our simulation settings, the null hypothesis—that the ATS was equal across two groups—was tested.
We employed four distinct estimators for this purpose: the MC, CS, ANCOVA, and the slope of a straight line fit, each assessed at a significance level of $\alpha=0.05$.
The evaluation of power was based on the mean trajectories depicted in the left panel of Figure \ref{f0}, while the Type I error assessment utilized the scenario presented in the right panel of Figure \ref{f0}.

\begin{figure}[htp]
\caption{\textbf{Power Analysis for the Quadratic Trajectories}}
\centering
\includegraphics[width=1\textwidth]{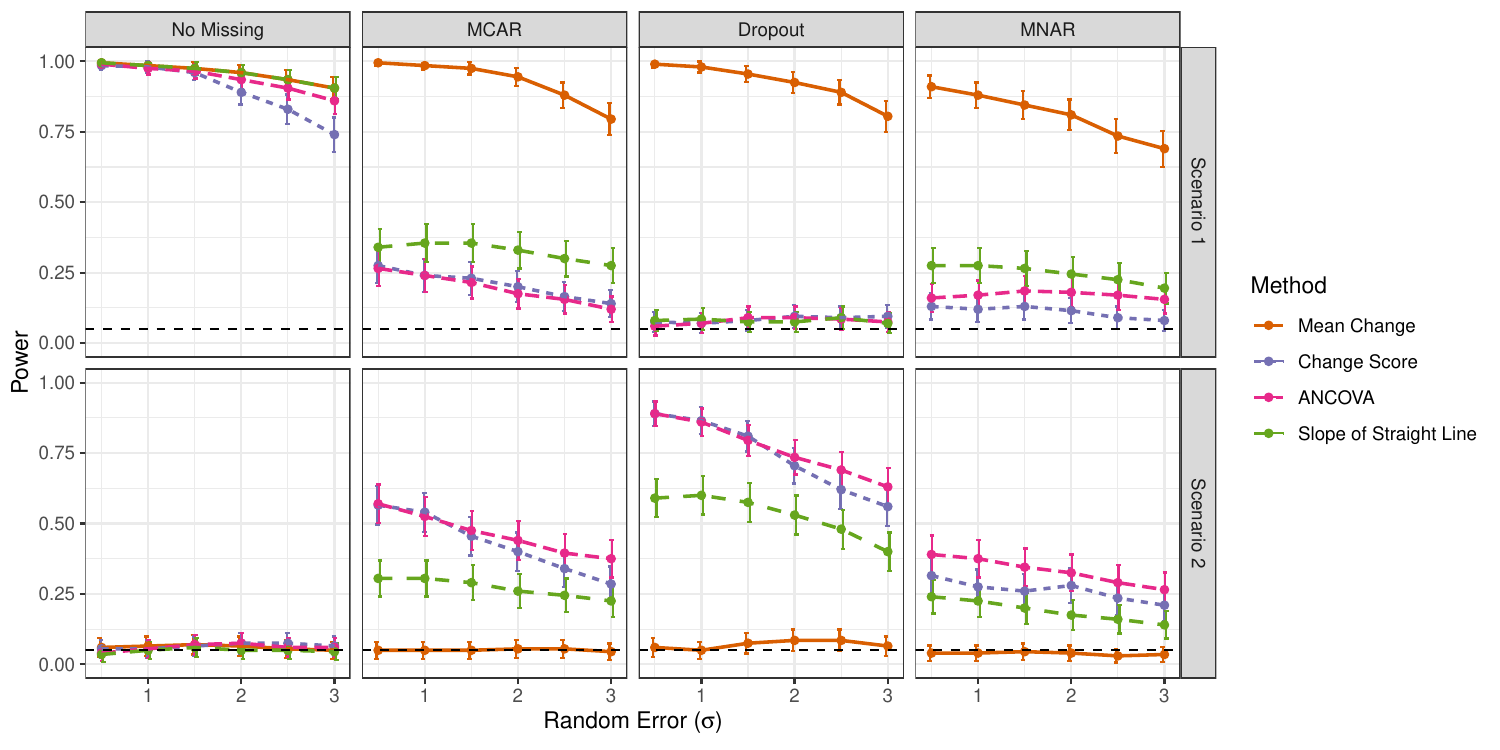}
\label{fq}
\caption*{
\footnotesize{
The points in each panel represent the mean proportion of null hypothesis rejections across 1000 simulations and the bars represent the standard deviations of the rejection proportion.
The results in the top row represent the Scenario 1, when the $H_0$ is false (i.e., power); 
the results in the bottom row represent the Scenario 2, when the $H_0$ is true (i.e., Type I error).
}}
\end{figure}

\begin{figure}[htp]
\caption{\textbf{Power Analysis for the Quadratic Trajectories with MI}}
\centering
\includegraphics[width=1\textwidth]{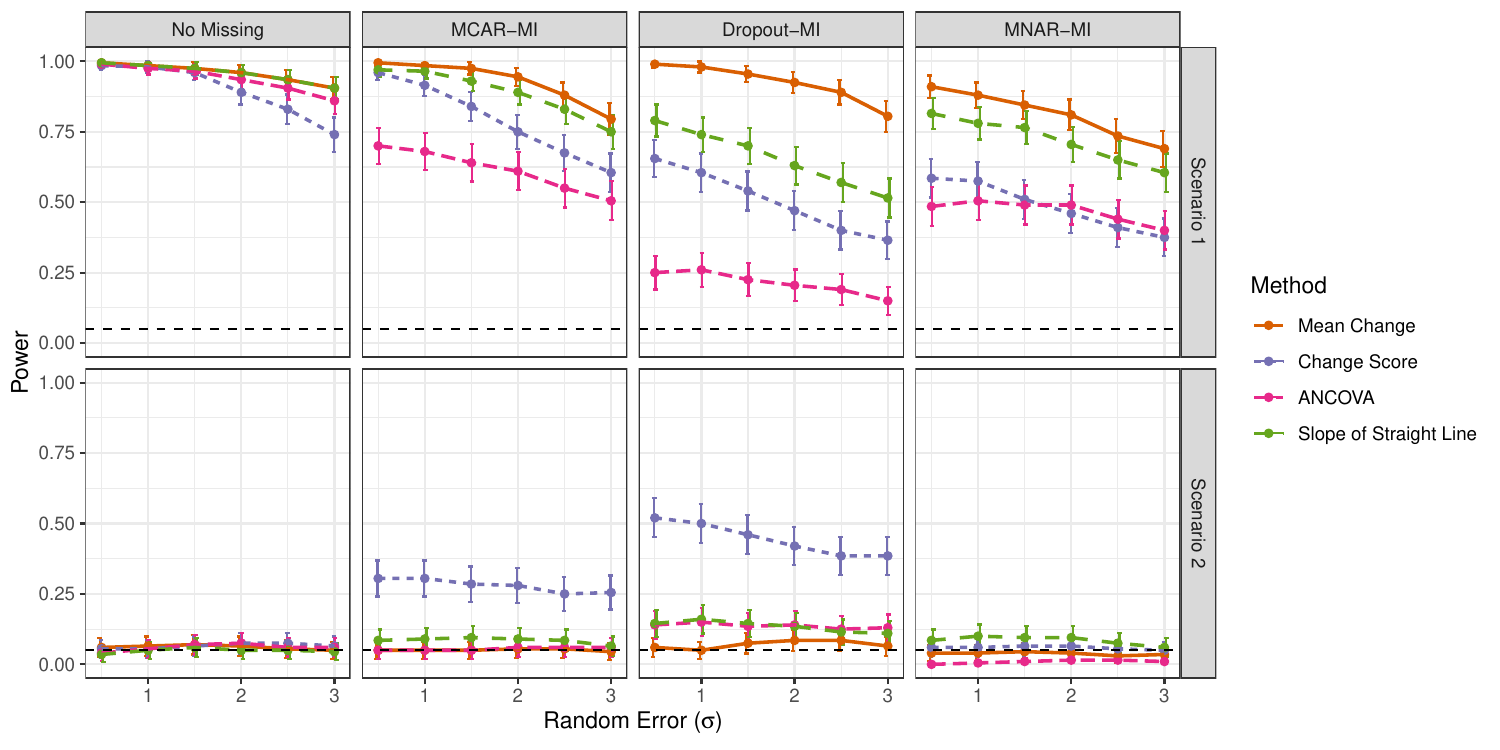}
\label{fqmis}
\caption*{
\footnotesize{
MI was conducted for the CS, ANCOVA, and slope of the straight line to handle missing values.  
MC was estimated from the original available data.
Missingness scenarios across columns and rows correspond to different mean trajectory shapes shown in Figure \ref{f0}.
The points in each panel represent the mean proportion of null hypothesis rejections across 1000 simulations and the bars represent the standard deviations of the rejection proportion. 
The results in the top row correspond to power (Scenario 1, when the $H_0$ is false);
the bottom row shows the type I error (Scenario 2, when the $H_0$ is true).}}
\end{figure}

Figure \ref{fq} shows the power and type I error for the quadratic trajectory scenarios 1 and 2 and 
Figure \ref{fqmis} shows results when MI is used for missing data (plots for scenario 3 are not shown).
Within these figures, the columns, from left to right, correspond to the four missing data scenarios: no missing data, MCAR, dropout, and MNAR. 
The rows, from top to bottom, represent the results for Scenarios 1 (non-null) and 2 (null), as depicted in Figure \ref{f0}. 
The $x$-axis in each panel indicates the level of noise,  $\sigma$, and a horizontal dashed line marks the significance level (i.e., $\alpha = 0.05$).
The top row of Figure \ref{fq} demonstrates a decrease in power for all methods as the random error $\sigma$ increases, aligning with expectations.
In the case of the MC, results presented in Figure \ref{fqmis} did not incorporate MI. 
This decision was based on observations that MI tended to adversely impact the MC estimator's performance, likely by inflating the variance of the estimator.
The following points summarize the simulation results:
\begin{itemize}
\item
The type I error rate was inflated using the CS, ANCOVA, and the slope of a straight line approaches for all missing data settings
(columns 2--4 in Figure (\ref{fq})) except for scenario 3 when the mean trajectories are identical for both groups,
whereas the MC estimator protected the type I error rate across all scenarios and for each missing data scenario.
\item
The inflation of type I error was significantly reduced with MI, as depicted in Figure \ref{fqmis}. 
However, the CS method continued to exhibit a substantial elevation in type I error, even with the application of MI.
\item
When there was no missing data, the MC and the slope of fitting a straight line yielded the highest power. 
Note that the straight-line slope fitting provides an unbiased estimation of the ATS in quadratic trajectory contexts when data is fully present \citep{tarpey2021extracting}.
\item
Only the MC method maintained high power for each missing data setting (columns 2--4 in Figure \ref{fq}).
Other methods, particularly in dropout scenarios, experienced a significant reduction in power. 
This is likely attributed to the crossover of mean trajectories between week 6 and 7 in Figure \ref{f0} and the prevalence of dropouts at week 6, which diminished the differences between the two groups.
\item
Comparing Figures \ref{fq} and \ref{fqmis},
MI tended to provide a marginal improvement in power for the CS, ANCOVA and the slope of a straight line estimator, but even with MI, each of these methods were still inferior to the MC estimator.
\item
MI tended to improve the performance of the CS method more so compared to the ANCOVA estimator in terms of power,
(top row of Figure \ref{fqmis})
but the CS did generally worse in protecting the type I error rate.
It is interesting to note that 
the type I errors of these three approaches were still inflated in some missingness settings after MI (e.g. MCAR and dropout in Scenario 2) whereas the type I error rate was controlled using the MC estimator.
\end{itemize}

In summary, across various scenarios, the MC estimator consistently outperforms CS and ANCOVA in terms of power and Type I error protection, a trend that holds even in the presence of missing data addressed with MI.
Also note that CS and ANCOVA do not consistently follow a trend of accepting the null hypothesis. 


Instead, they appear to yield counterintuitive results, such as incorrectly rejecting the null when it is true, or failing to reject the null when it is false.
This pattern suggests that CS and ANCOVA may not be robust in certain missing data scenarios, particularly with dropout.
Additionally, we extended our simulations to include various mean curve shapes beyond the quadratic, employing $B$-splines for the MC estimator in scenarios where the mean curve shape was unspecified. 
The findings from these additional simulations align closely with those reported here (please see the Appendix for detailed results on these non-quadratic simulations). 

\section{Choosing a Weight Function for WATS} \label{watsection}

The MC estimator of the ATS efficiently extracts a scalar summary from a curve as a special case of the weighted average tangent slope (WATS) (\ref{wats0}) using a uniform weight function $w(t)$.
This section describes an empirical approach to estimate a flexible weight function for the WATS in order to gain additional insight in order to distinguish outcome trajectories across groups.
From (\ref{wats0}), let ${\theta}_{w,k}  
  =    \int_{t_1^*}^{t_m^*} w(t) \bm g^\prime(t)\tran \bm \beta_k dt = \bm S_w \tran \bm \beta_k,$
where 
$\bm S_w = \Big(
\int_{t_1^*}^{t_m^*} w(t) g_1^\prime(t)dt, ..., 
\int_{t_1^*}^{t_m^*} w(t) g_p^\prime(t)dt\Big) \tran$.
The criterion considered for the weight function $w(t)$ is to maximize the squared standard distance between the weighted MC estimators $\reallywidehat{\theta}_{w,k}$.
From 
(\ref{wats0}),
covariance matrix of
$\reallywidehat{\text{Var}} (\reallywidehat{\theta}_{w,k})$ is $\bm S_w \tran \reallywidehat{\text{Cov}} (\reallywidehat{\bm \beta}_k)  \bm S_w$
and therefore the criterion for estimating the weight function for the WATS is given by
\begin{equation} \label{7}
\begin{aligned}
{\reallywidehat w(t)} = 
 \argmax_{w(t)\ge 0} \frac{ \bm S_w \tran (\reallywidehat{\bm \beta}_1 - \reallywidehat{\bm \beta}_2) (\reallywidehat{\bm \beta}_1 - \reallywidehat{\bm \beta}_2) \tran \bm S_w }{ \bm S_w \tran \big( \reallywidehat{\text{Cov}} (\reallywidehat{\bm \beta}_1) + \reallywidehat{\text{Cov}} (\reallywidehat{\bm \beta}_2) \big) \bm S_w }.
\end{aligned}
\end{equation}
The optimal weight function can be normalized to integrate to one by multiplying by a scalar.  
To enforce the non-negative constraint on $w(t)$, we implemented the approach in
\cite{chen2014optimally} by considering
\begin{equation}\label{watsgen}
w(t) = [\bm u_w\tran (t) \bm v]^,
\end{equation}
where $\bm u_w (t) = [u_{w1} (t), ... u_{wk_w} (t)]\tran$ denotes a vector of spline basis functions and $\bm v$ is the vector of associated coefficients. 
Note that, the basis function for the mean trajectory, $\bm g(t)$ and the basis function for the weight, $\bm u_w(t)$ can be different. 
The WATS is then defined as:
\begin{equation}\label{watsgen2}
\begin{aligned}
\theta_{w,k} = \int w(t) \bm g^\prime(t) \tran \bm \beta_k dt = \bm v \tran \bm M_{w, \bm \beta_k} \bm v  = \bm v \tran \bm H_{w, \bm v} \bm \beta_k,
\end{aligned}
\end{equation}
where 
$$\bm M_{w, \bm \beta_k}
= \int \bm u_w(t)\bm u_w \tran (t)\bm g^\prime (t) \tran \bm \beta_k dt \;\;\text{and}\;\;\;
\bm H_{w, \bm v} = \int  \bm u_w(t) \bm u_w\tran (t) \bm v \bm g^\prime (t) \tran  dt.$$ 
From (\ref{watsgen2}), the square difference between the estimated WATS between two groups is: 
\begin{equation}
\begin{aligned}
(\reallywidehat \theta_{w,1} - \reallywidehat\theta_{w,2})^2= 
\bm v\tran \reallywidehat{\bm A}_{w, \bm v, \bm \beta_k} \bm v,
\end{aligned}
\end{equation}
where $\reallywidehat{\bm A}_{w, \bm v, \bm \beta_k} = (\reallywidehat{\bm M}_{w, \bm \beta_1}  - \reallywidehat{\bm M}_{w, \bm \beta_2} )\bm v \bm v \tran  (\reallywidehat{\bm M}_{w, \bm \beta_1} -\reallywidehat{\bm M}_{w, \bm \beta_2} )\tran$
and the variance of $\reallywidehat{\theta}_{w,1} - \reallywidehat{\theta}_{w,2}$ is 
\begin{equation}
\begin{aligned}
\reallywidehat{\text{Var}}(\reallywidehat{\theta}_{w,1} - \reallywidehat{\theta}_{w,2}) = 
\bm v \tran \reallywidehat{\bm B}_{w, \bm v, \bm \beta_k} \bm v
\end{aligned}
\end{equation}
where $\reallywidehat{\bm B}_{w, \bm v, \bm \beta_k} = \bm H_{w, \bm v} \Big(\reallywidehat{\text{Cov}} (\reallywidehat{\bm \beta}_1)  + \reallywidehat{\text{Cov}} (\reallywidehat{\bm \beta}_2) \Big)\bm H_{w, \bm v} \tran$. 
Thus the coefficients defining the optimal weight function are obtained from
\begin{equation}\label{watsgen3}
\reallywidehat{\bm v} = \argmax_{\bm v} \frac{\bm v\tran \reallywidehat{\bm A}_{w, \bm v, \bm \beta_k} \bm v}{\bm v\tran \reallywidehat{\bm B}_{w, \bm v, \bm \beta_k} \bm v}.
\end{equation}
The matrices $\reallywidehat{\bm A}_{w, \bm v, \bm \beta_k}$ and $\reallywidehat{\bm B}_{w, \bm v, \bm \beta_k}$ depend on $\bm{v}$ and hence an iterative approach is needed to maximize the Rayleigh quotient
(\ref{watsgen3}) and we implemented the Nelder Mead Algorithm \citep{nelder1965simplex} to obtain the estimator $\reallywidehat{\bm v}$ in (\ref{watsgen3}).
Parametric weight functions based on an exponential function and a beta probability density function were also considered, but the weight function proposed here provided multi-modal flexibility.

The power and type I error rates for the weighted MC were evaluated using simulation scenarios described in Section \ref{simulation1} (results not shown).
Although the power for the weighted MC was much higher compared to the other ATS estimation procedures, the weighted MC tended to inflate the type I error rate considerably which is expected since the data-driven estimated weight function is estimated precisely to highlight differences in the mean trajectories between groups.

\section{Application to Depression Data} \label{embarc}

\begin{figure}[htp]
\caption{\textbf{Outcome Trajectories Plots for Participants in the EMBARC Study}}
\centering
\includegraphics[width=0.5\textwidth]{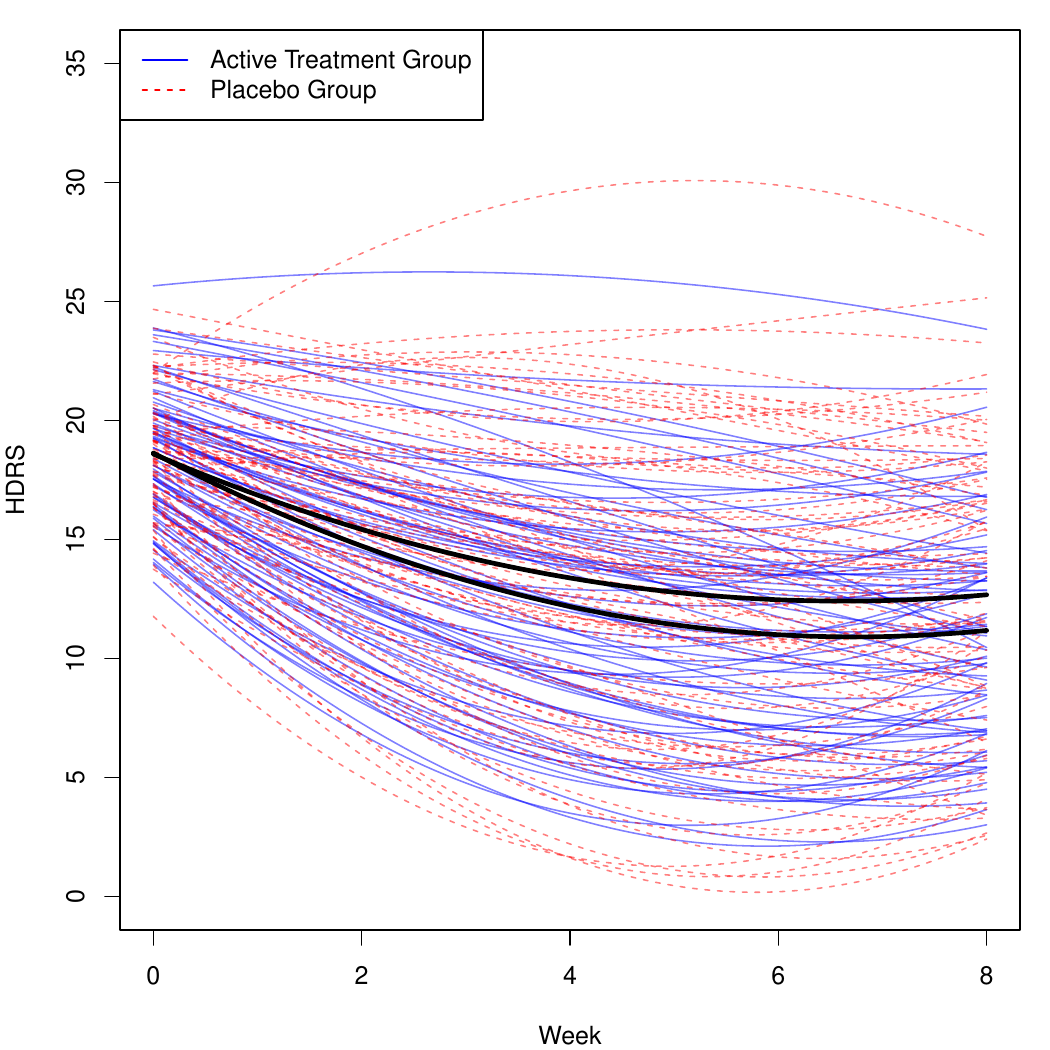}
\label{fig:fig1}
\caption*{
\footnotesize{
Predicted quadratic trajectories for the HDRS outcomes obtained from fitting linear mixed-effect models\thad{.}
The blue solid and red dashed curves represent the active treatment and placebo groups' estimated outcome trajectories, respectively. 
The average trajectories for the two groups are shown by the two bold black curves (the active treatment mean curve is the lower curve).}}
\end{figure}

To illustrate the WATS and ATS estimation approaches, we use
data from the antidepressant randomized clinical trial ``Establishing Moderators and Biosignatures of Antidepressant Response in Clinical Care'' (EMBARC) \citep{trivedi2016establishing} 
where participants were randomized 50-50 to receive either sertraline or placebo. The primary outcome, Hamilton Depression Rating Scale (HDRS), was evaluated for each participant at weeks 0, 1, 2, 3, 4, 6, and 8. 
The HDRS is a measure for the severity of depression with lower scores indicating less severe depression.
287 participants had at least two observed outcomes (including baseline): 
  64$\%$ of participants had all assessments and
$17$\%,  $7$\%, $5$\%, $2$\%, $2$\%, and $3$\% of the participants 
were missing one, two, three, four, five, and six assessments, respectively. 
The rate of missingness was distributed relatively equally between the two intervention groups.
Mixed-effect\st{s} models  with quadratic time trends were fit to the longitudinal data and the fitted trajectories are shown in Figure \ref{fig:fig1} (blue solid and red dashed curves represent the estimated trajectories for the active treatment and placebo group, respectively). 
The mean outcome trajectories for the two treatment groups are represented by the thick black curves (the mean trajectory for the active treatment group is the lower one). 
Note that the average trajectories for the treatment and placebo are very similar to one another.

\subsection{One-Sided Test to Compare Active to Placebo}\label{1side}

A likelihood ratio test (LRT) based on the mixed-effects model to test if outcome distributions differed for active versus placebo produced a $p$-value $=0.036$ (degree of freedom = 10). 
Although the LRT indicates that the trajectory distributions for active and placebo differ, the test does not
indicate in what respect the distributions differ, e.g., do they differ in terms of fixed-effects, the error variances, and/or the random effect distributions? In particular, the LRT does not answer the question of whether  participants treated with the active drug have better outcomes on average compared to placebo-treated individuals.  
However,
the MC, CS, and ANCOVA testing approaches can be used for one-sided inference to test if there is more improvement on the active drug compared to placebo.
The one-sided $p$-values for the CS and ANCOVA approaches were $p=0.192$ and $0.146$ respectively.
The MC approach however yielded a one-sided $p=0.055$ using the Wald test in Section \ref{sectionats}.
Therefore, the testing based on the MC indicated modest evidence of a more improvement on the active drug compared to placebo on average, whereas the CS and ANCOVA testing approaches lack the power to detect this evidence
of improvement.

\subsection{Estimating a Flexible Weight Function}\label{flex}

\begin{figure}[htp]
\caption{\textbf{Weight Function Estimated from the EMBARC Data}}
\centering
\includegraphics[width=0.5\textwidth]{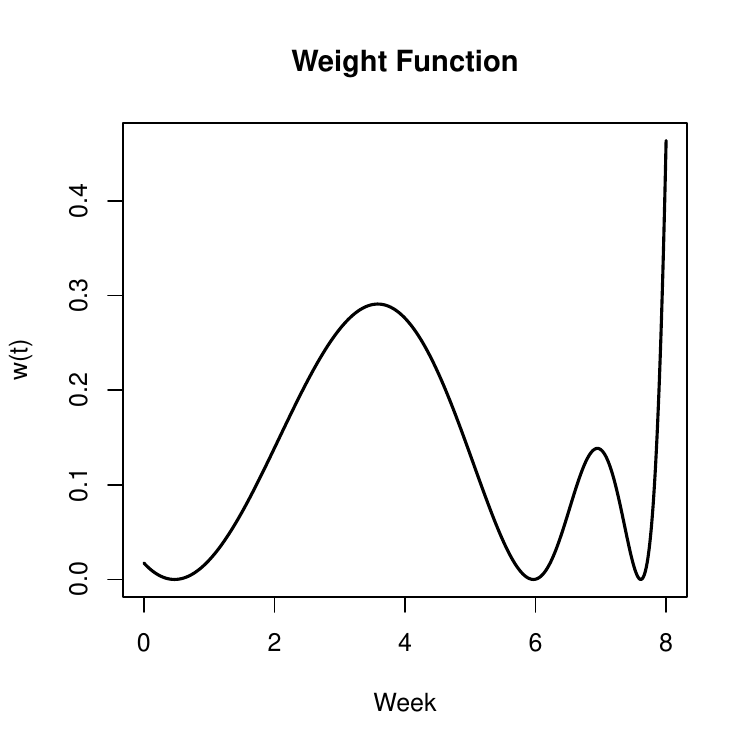} 
\label{density}
\caption*{
\footnotesize{
Estimated weight function for the EMBARC data set with quadratic mixed-effect models are shown. The parameters of the weight function were optimized with the methods in Section \ref{watsection}.}}
\end{figure}

To highlight time periods during the trial where the average rate of improvement in the active treatment arm is differentiated from the placebo response seen in the control arm,
the spline-based WATS weight function 
(\ref{watsgen}) was estimated and is shown in Figure \ref{density}.
Unsurprisingly perhaps, the weight function emphasizes the latter portion of the treatment period where the mean trajectories show greater divergence; interestingly, the weight function also shows a mode around week 4, indicating
that the greater rate of improvement for the active treatment compared to the placebo around the halfway point
helps distinguish the two groups in terms of the rate of improvement.
By emphasizing specific periods during the study, the estimated weight function for the WATS highlights the differences between the active treatment and the placebo group 
trajectories, thus making it easier to identify periods of differential treatment effects. 
Additional details on the WATS for this example are provided in the Appendix.

\section{Discussion} \label{discussion}

It is very common for randomized studies (e.g., RCTs) to collect multiple outcomes longitudinally which can allow the modeling of change in the outcome over time.
However, the 
standard approaches used to compare groups in these settings (Change Score (CS) and ANCOVA) use only the first and last observations and are inherently inefficient. 
We have shown that the Mean Change (MC) estimator, that capitalizes on the information from the longitudinal modeling, performs substantially better the CS and ANCOVA estimators 
in terms of power and control of type I error rates. This superior performance becomes more enhanced when missing data is an issue.
A potential weakness of the MC estimator is its dependency on the model specification.
However, for most randomized experiments, the time durations of studies tend to be fairly short where simple models (quadratic or cubic-spline with a single knot)
often produce good fits to the data and, as we have shown that, the MC estimator performs well in these scenarios.
A useful extension to explore as a follow up to the results presented here would be to explore the efficiencies of the MC estimation using nonparametric inference when flexible smoothing approaches are used to extract the fitted outcome trajectories \citep[e.g.,][]{Silverman1985, Wood2017}.

A scenario where the performance of the MC estimator may degrade is when the outcome trajectories approach an equilibrium plateau (horizontal asymptote) as might happen for longer studies. In such cases, polynomial-based models (e.g., splines) may fail to capture this characteristic whereas the CS may be robust to this feature, especially if dropout is an issue and participants tend to drop out after the plateau is reached.  However, if a correctly specified nonlinear regression model is available, then the MC estimator using nonlinear regression would likely perform superiorly to the CS and ANOVA approaches.

This paper has primarily focused on estimators of the ATS for inference when comparing groups.  
Another motivation for this paper is to determine approaches to extract scalar summaries from outcome curves to develop powerful individualized treatment rules (ITRs) in precision medicine. Initial results reported by \cite{yao2022single} show that defining ITRs that incorporate information from the entire outcome trajectory produces ITRs that perform better than currently available ITR approaches that use only scalar outcomes.  We anticipate that the use of
a weighting function as in the WATS to highlight differences in outcome trajectories between treatment groups will lead to powerful ITRs for
advancing precision medicine research.

\section*{Acknowledgements}
This work is supported by the National Institute of Mental Health (NIMH), Grant/Award Number: 5 R01 MH099003.

\newpage

\appendix
\counterwithin{figure}{section}

\begin{center}
    \begin{Large}
  \textbf{Appendix}
    \end{Large}
\end{center}

\section{Simulation Details} \label{aaaa}

When the trajectories are quadratic, we set $\bm X_{ik} = (\bm 1, \bm t_{ik}, \bm t_{ik}^2)$. The fixed effects $\bm \beta_k$ were set as 
\begin{equation}
    \begin{aligned}
  & \text{Quadratic Group 1 and Group 4: }  \bm \beta_1 = (20, -2, 0.2)\tran,  \\
  & \text{Quadratic Group 2: }  \bm \beta_2 =  (20, 1.2, -0.3)\tran, \\
  & \text{Quadratic Group 3: }  \bm \beta_3 = (20 - 4.8, 0.6)\tran,
    \end{aligned}
\end{equation}
with mean trajectories:
\begin{equation}
    \begin{aligned}
  & \text{Quadratic Group 1 and Group 4: }  20 - 2 t + 0.2 t^2,  \\
  & \text{Quadratic Group 2: }  20 + 1.2 t - 0.3 t^2,  \\
  & \text{Quadratic Group 3: }  20 - 4.8t + 0.6 t^2.
    \end{aligned}
\end{equation}
The random effects are $\bm b_{ik} \sim MVN (\bm 0, \bm D_k)$, where $\bm D_k$ is set as (note we set $\bm D_1 = \bm D_2$):
$$\bm D_k = \begin{pmatrix}
8 & 3 & -0.4 \\
3 & 1.5, & -0.16 \\
-0.4 & -0.16 & 0.03
\end{pmatrix}.$$
Non-quadratic simulations generated random effects with the same covariance matrix $\bm D_1$ and $\bm D_2$.

\section{Simulation Results for Non-Quadratic Trajectories}\label{nonquad2_}

In this section, we present an in-depth analysis of the simulation results focusing on power and type I error rates in the context of ATS tests, particularly when dealing with complex-shaped, non-quadratic trajectories. 

\subsection{Simulation Settings} \label{simsetting}

\begin{figure}[H]
\caption{\textbf{Simulated Outcome Trajectories}}
\centering
\includegraphics[width=0.9\textwidth]{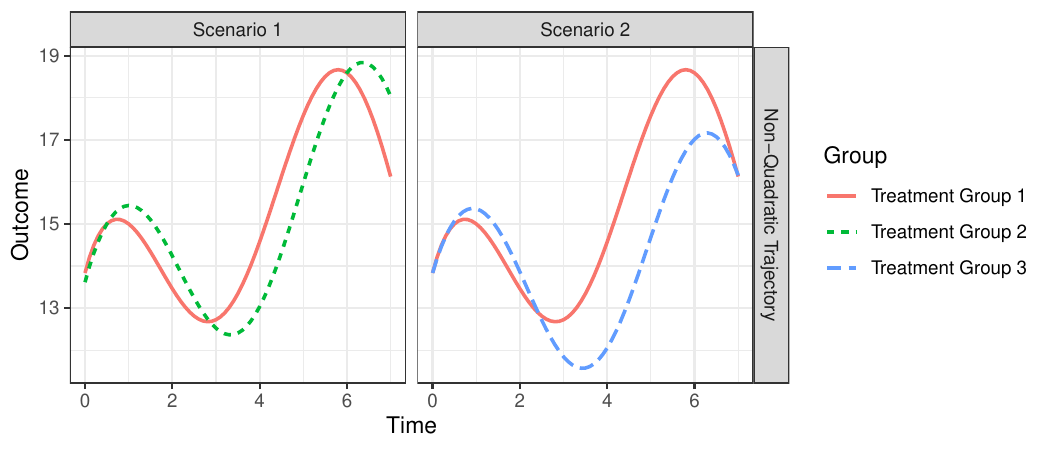}
\label{fignonquad}
\caption*{
\footnotesize{
The mean trajectories for various groups under non-quadratic scenarios are shown.
Scenario 1: the two mean curves have different shapes and different ATSs; Scenario 2: the two mean curves have different shapes but the same ATS.
For Scenario 3 (not shown), the two mean trajectories are identical for both groups (e.g. the red curve shown in the right panel).}}
\end{figure}

We delve into a variety of curve shapes through the simulation of distinct mean trajectory scenarios, as illustrated in Figure \ref{fignonquad}.
These trajectories are non-quadratic, expanding upon the quadratic scenarios previously discussed in Section \label{hyptest}.
Our analysis focuses on three specific scenarios of mean trajectories:
(i) the ATSs differ for the two groups, 
(ii) the ATSs are the same for the 2 groups but mean trajectories differ, 
and (iii) mean trajectories are identical for both groups.
In every scenario, the mean baseline value is consistent for both groups, aligning with expectations from randomized experiments. 
Scenarios 2 and 3 simulate the null hypothesis, where the ATS is equivalent for both groups. 
However, in Scenario 2, the mean trajectories diverge despite having the same ATS.
The non-quadratic mean trajectories shown in the bottom row of Figure \ref{fignonquad} were generated using the following equations:
\begin{equation}\label{nonquadsim}
    \begin{aligned}
    & \text{Group 1 and 4: } 15 - 2\sin(t-1) \log(t+0.5), \\
    & \text{Group 2: } 15 + 2\cos(t) \log(t+0.5), \\
    & \text{Group 3: } 15.22 -0.3 t + 2 \cos(t)\log(t+0.5).
    \end{aligned}
\end{equation}
The settings for missing data and error variances have been kept identical to those used in the quadratic simulation study as described in Section \ref{hyptest2}.
To implement the estimation of the ATS using MC, 
Cubic $B$-splines (with a single knot point at the half-way time point) were used to estimate the models with non-quadratic mean trajectories in each group separately using random effects for the intercept, linear, and quadratic terms only (model fits with additional random effects were unstable).
For the fitting of cubit $B$-spline models, the R package \code{splines} was employed, while the package \code{mice} was utilize for MI. 
In summary, the simulation study evaluated the following combinations of mean trajectory settings shown in the panels of Figure \ref{fignonquad}).
Under each scenario, a sample size of $n_k=100$ was generated for each group and the CS, MC, ANCOVA as well as the slope of a straight line fit were used to estimate the ATS.

\subsection{Simulation Results} \label{simresult2}

\begin{figure}[htp]
\caption{\textbf{Power Analysis for the Non-Quadratic Trajectories}}
\centering
\includegraphics[width=1\textwidth]{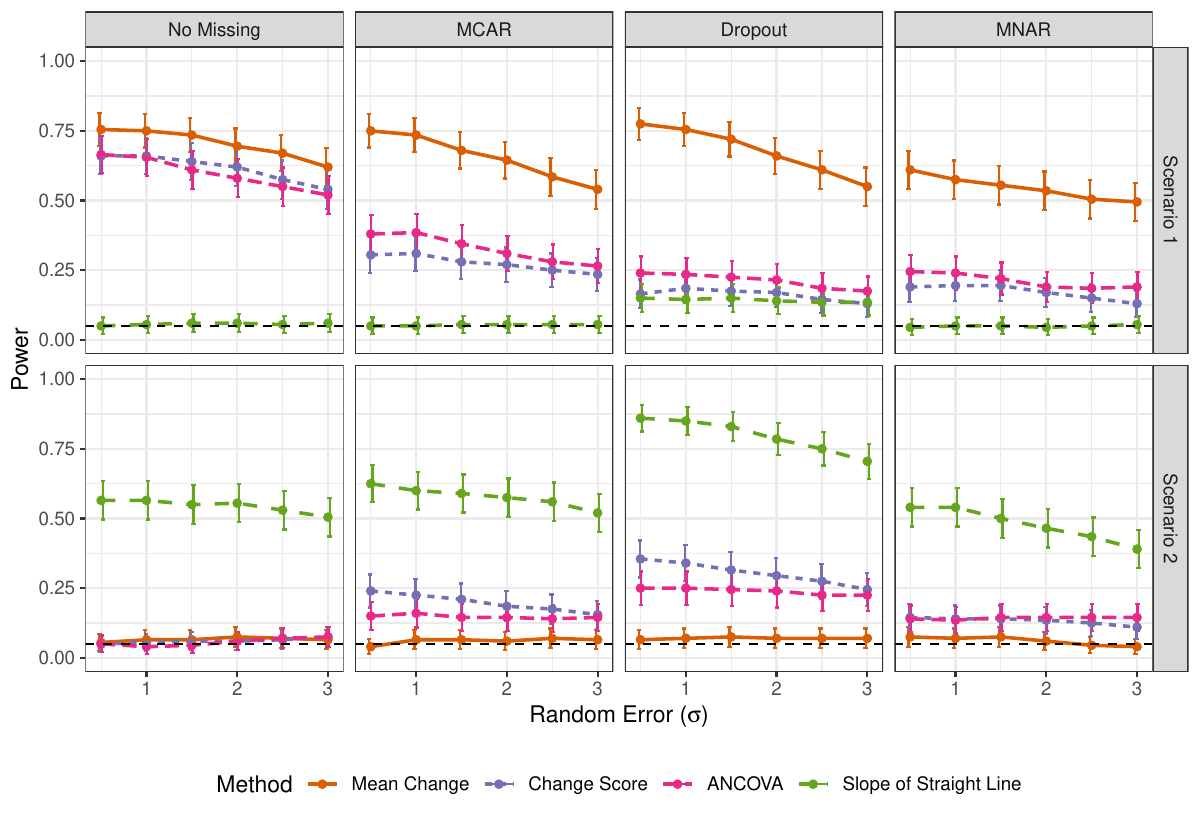}
\label{fapp_non_quad}
\caption*{
\footnotesize{
The points in each panel represent the mean proportion of null hypothesis rejections across 1000 simulations and the bars represent the standard deviations of the rejection proportion.
The results in the top row represent the Scenario 1, when the $H_0$ is false (i.e., power); 
the results in the bottom row represent the Scenario 2, when the $H_0$ is true (i.e., Type I error).
}}
\end{figure}

\begin{figure}[htp]
\caption{\textbf{Power Analysis for the Non-Quadratic Trajectories with MI}}
\centering
\includegraphics[width=1\textwidth]{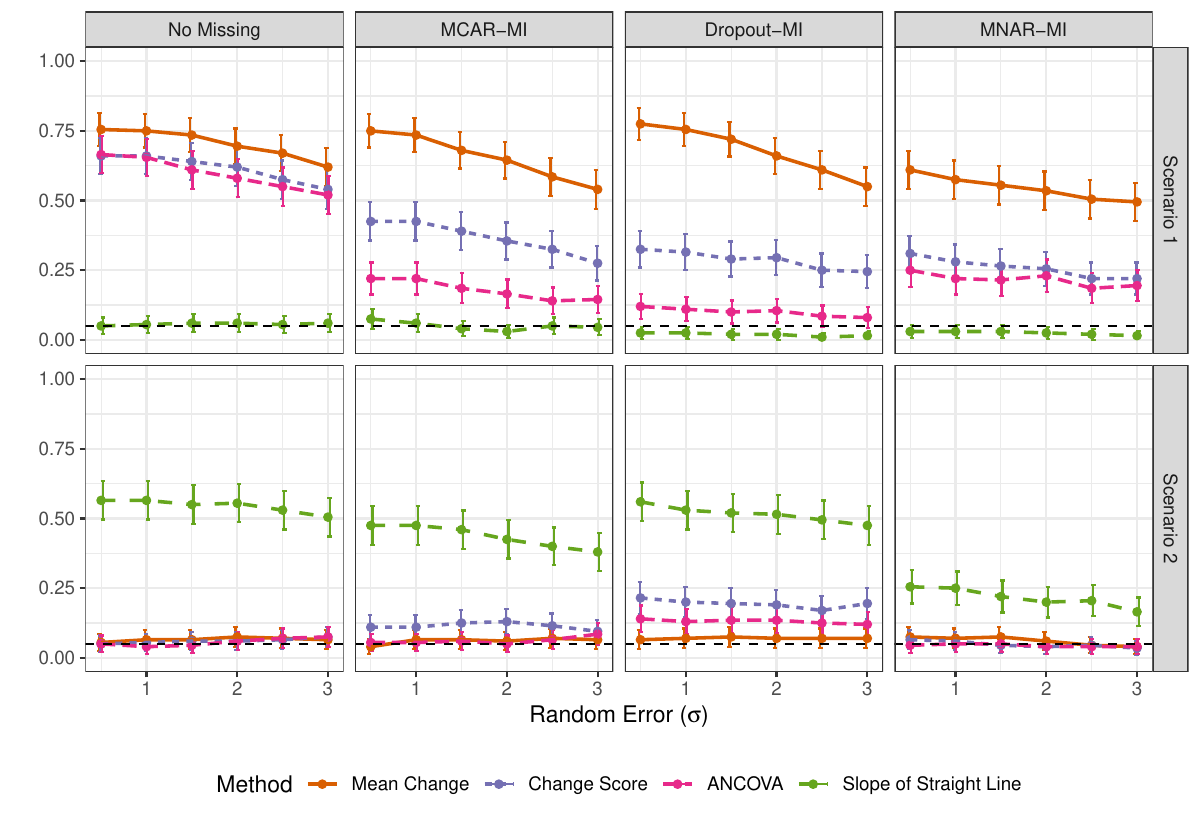}
\label{fapp_non_quad_mi}
\caption*{
\footnotesize{
MI was conducted for the CS, ANCOVA, and slope of a straight line to handle missing values.  
Mean Change was estimated from available data.
Missingness scenarios across columns and rows correspond to different mean trajectory shapes shown in Figure \ref{fignonquad}.
The points in each panel represent the mean proportion of null hypothesis rejections across 1000 simulations and the bars represent the standard deviations of the rejection proportion. 
The results in the top row correspond to power (Scenario 1, when the $H_0$ is false);
the bottom row shows the Type I error (Scenario 2, when the $H_0$ is true).}}
\end{figure}

For all simulation settings, we assessed the null hypothesis, which posits equal ATS for two groups, using a range of methods including MC, CS, ANCOVA, and slope of fitting a straight line. 
These methods were applied at a significance threshold of $\alpha = 0.05$. 
To evaluate the power of these tests, we referred to the mean trajectories illustrated in the left panel of Figure \ref{fignonquad}. 
For the assessment of Type I error, we focused on the results from Scenario 2, as presented in the right panel of Figure \ref{fignonquad}.
Under the non-quadratic trajectory setting, Figure \ref{fapp_non_quad} illustrates the power and Type I error rates for scenarios 1 and 2, while Figure \ref{fapp_non_quad_mi} presents the outcomes when MI is applied to handle missing data (plots for scenario 3 are not shown).

In the extended simulation study focusing on non-quadratic trajectories, our findings echoed those observed in the quadratic scenario, as delineated in Section \ref{hyptest2}.
However, the key distinction observed in the non-quadratic setting was the underperformance of the slope of fitting a straight line method, which persisted even after applying.
Unlike in quadratic trajectories, the slope method demonstrated a consistently low power of around 0.05 in non-quadratic contexts, coupled with an inflated Type I error rate, particularly noticeable in scenario 2. 
This suggests that the effectiveness of the slope method is restricted to quadratic trajectories and its application in non-quadratic settings may lead to suboptimal outcomes, especially in the presence of missing data.

In summary, the non-quadratic simulation results reinforce the superiority of the MC estimator in both power and Type I error protection across various scenarios, including those with missing data treated with MI. This trend is consistent with findings from the quadratic simulations. 
However, methods like CS and ANCOVA do not show a consistent trend in hypothesis testing, often yielding counterintuitive outcomes such as false rejection or failure to reject the null hypothesis under certain conditions, particularly in dropout scenarios. 
This indicates potential limitations in the robustness of CS and ANCOVA in specific missing data contexts.

\section{Additional Details on the WATS for the EMBARC Example}\label{EMBARCadditional}

\begin{figure}[htp]
\caption{\textbf{Density Plots of the Summary Statistics from the EMBARC Data}}
\centering
\includegraphics[width=\textwidth]{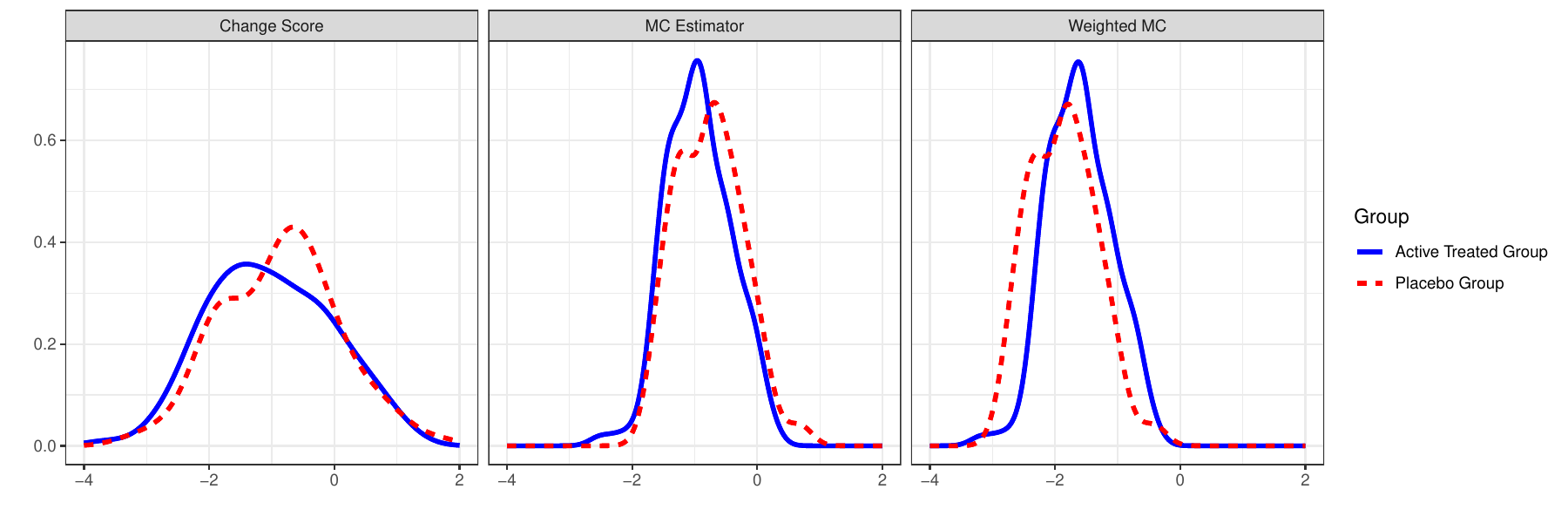}
\label{denplots}
\caption*{
\footnotesize{Density plots of the change scores, MC, and weighted MC calculated for the subjects in the EMBARC study. 
}}
\end{figure}

The estimated weight function depicted in Figure \ref{watsection} was employed to compute the weighted MC for each participant in the EMBARC study.
With the estimated weight function, $\reallywidehat w(t)$, we can determine the design matrix $\reallywidehat{\bm S}_w$, which is computed as 
$\Big(
\int_{t_1^*}^{t_m^*} \reallywidehat w(t) g_1^\prime(t)dt, \ldots, 
\int_{t_1^*}^{t_m^*} \reallywidehat  w(t) g_p^\prime(t)dt\Big) \tran$. 
Utilizing the Best Linear Unbiased Predictor (BLUP) approach, the individual-specific effects, denoted as $\reallywidehat{\bm \beta}_{ki}$, for each participant can be estimated. 
Therefore, following the method outlined in (\ref{wats0}), the individualized WATS can be estimated as $\reallywidehat{\bm S}_w \tran \reallywidehat{\bm \beta}_{ki}$.
Likewise, the MC and CS estimators were calculated for each subject. Figure \ref{denplots} presents the density distributions for the CS, MC, and weighted MC for both the active (blue) and control (red) treatment groups. Notably, the MCs and weighted MCs exhibit similar standard deviations across both treatment groups. 
In contrast, the variability observed in the CS is significantly larger than that of the other two measures. 
Consequently, the use of MC and weighted MC may offer higher statistical power compared to the CS in this dataset.

\begin{figure}[htp]
\caption{\textbf{Illustrations with Estimated Trajectories.}}
\centering
\includegraphics[width=0.9\textwidth]{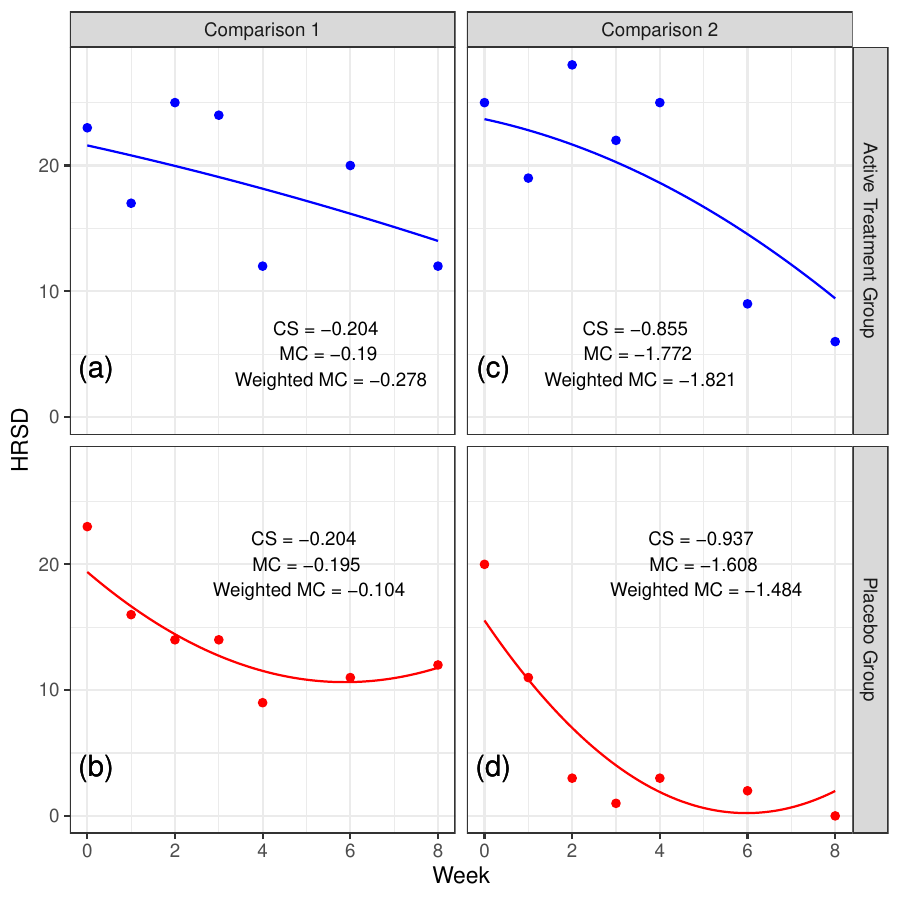}
\label{comparison}
\caption*{
\footnotesize{
Illustration of estimated trajectories.
Outcomes of four subjects in the EMBARC data set. 
The subjects shown in the upper and lower rows are from the active treatment group and the placebo group, respectively. 
The values of CS, MC, and weighted estimated MC are labeled in each panel. 
Dots: observed data; Curves: estimated quadratic trajectories. 
}}
\end{figure}
Figure \ref{comparison} displays outcomes from four participants from the EMBARC data set,  alongside their corresponding estimated trajectories.
The top row features subjects from the active treatment group, while the bottom row includes those from the placebo group.
Solid points on the graphs represent observed outcomes for each individual across the study weeks.
These graphs also show fitted quadratic trajectories, which are the results of separately fitting linear mixed-effect models to each treatment group.
The models have both linear and quadratic fixed and random effects.
The panel texts provide the CS, the MC, and the weighted MC estimators in terms of $z$-scores since the scales for ATS and the WATS are not the same.

Despite having identical Change Scores (CS), patients in panels (a) and (b) exhibit markedly different response patterns, as revealed by their scatter plots. 
The depressive symptoms of the subject in panel (a) showed a consistent deterioration, whereas the patient in panel (b) initially experienced an improvement in symptoms, followed by a subsequent worsening (a higher HRSD score indicates more severe symptoms). 
The CS estimator, which considers only the initial and final data points, fails to capture these nuanced differences in their trajectories.
The MC estimator detected a slight difference between these patients. 
As Figure \ref{density} shows, the weight function emphasizes the last observation and data around week 4. 
For trajectory (a), the tangent slopes remained negative, indicating a continuing downward trend. 
In contrast, trajectory (b) exhibited positive tangent slopes towards the end of the treatment, signaling a worsening condition. 
By assigning greater weight to the latter part of the observation period, the weighted MC for patient (b) was substantially higher (indicative of poorer health) than that for patient (a). 
This highlights the Weighted MC’s enhanced capability in distinguishing the distinct trajectory shapes between patients (a) and (b) compared to the MC estimator.

The participants shown in panels (c) and (d) exhibited greater improvement compared to those in panels (a) and (b).
Specifically, the CS, MC, and weighted MC for subjects in (c) and (d) were significantly lower than for those in (a) and (b).
Patient (c) showed a lower MC compared to (b), suggesting more substantial improvement for patient (c).
The trajectory in panel (c) is characterized by a downward concavity, whereas the curve in (d) displays an upward concavity. 
This indicates that patient (c) experienced initial symptomatic improvement, which accelerated over time. 
In contrast, subject (d) demonstrated a more rapid initial improvement than (c), but this was followed by a slowing and eventual deterioration towards the end of the trial.
It is important to recall that the weight function placed emphasis on data around weeks 4 and 8. 
Despite subject (d)'s more pronounced improvement in week 4, their rate of change turned positive between weeks 6-8. Consequently, the Weighted MC for (c) was much lower than for subject (d). 
The Weighted MC, with its focus on the trajectory's shape, effectively differentiates between patients based on the distinctive patterns of their outcome trajectories.

\end{document}